# Bulk Crystal Growth and Single-Crystal-to-Single-Crystal Phase Transitions in the Averievite CsClCu$_5$V$_2$O$_{10}$

Chao Liu,[1] Chao Ma,[1] Tieyan Chang,[2] Xiaoli Wang,[1] Chuanyan Fan,[1] Lu Han,[1] Feiyu Li,[1] Shanpeng Wang,[1] Yu-Sheng Chen,[2] and Junjie Zhang[1*]

[1]State Key Laboratory of Crystal Materials and Institute of Crystal Materials, Shandong University, Jinan 250100, Shandong, China

[2]NSF's ChemMatCARS, The University of Chicago, Lemont, IL 60439, United States

**ABSTRACT:** Quasi-two-dimensional averievites with triangle-kagome-triangle trilayers are of interest due to their rich structural and magnetic transitions and strong spin frustration that are expected to host quantum spin liquid ground state with suitable substitution or doping. Herein, we report growth of bulk single crystals of averievite CsClCu$_5$V$_2$O$_{10}$ with dimensions of several millimeters on edge in order to (1) address the open question whether the room temperature crystal structure is $P\bar{3}m1$, $P\bar{3}$, $P2_1/c$ or else, (2) to elucidate the nature of phase transitions, and (3) to study direction-dependent physical properties. Single-crystal-to-single-crystal structural transitions at ~305 K and ~127 K were observed in the averievite CsClCu$_5$V$_2$O$_{10}$ single crystals. The nature of the transition at ~305 K, which was reported as $P\bar{3}m1$-$P2_1/c$ transition, was found to be a structural transition from high temperature $P\bar{3}m1$ to low temperature $P\bar{3}$ by combining variable temperature synchrotron X-ray single crystal and high-resolution powder diffraction. In-plane and out-of-plane magnetic susceptibility and heat capacity measurements confirm a first-order transition at 305 K, a structural transition at 127 K and an antiferromagnetic transition at 24 K. These averievites are thus ideal model systems for a deeper understanding of structural transitions and magnetism.

1. Introduction

Geometrically frustrated systems including triangular, kagome, and honeycomb lattices have attracted much attention due to the emergence of exotic physical phenomena and many-body quantum states such as quantum spin liquid.[1-4] Among the materials studied, averievite, which represents a class of copper oxide minerals with the formula of (MX)$_n$Cu$_5$T$_2$O$_{10}$ (M=K, Rb, Cs, Cu; X=Cl, Br, I; n=1, 2…; T=P, V), are of current interest.[5-10] Specifically, averievite has a geometrically frustrated quasi-two-dimensional structure consisting of triangle-kagome-triangle trilayers of Cu$^{2+}$ (S=1/2).[7, 10-12] The ground state of averievites was found to be antiferromagnetic with Neel temperature varying from 24 K in CsClCu$_5$V$_2$O$_{10}$ (hereafter CCCVO)[7] to 3.8 K in CsClCu$_5$P$_2$O$_{10}$[8] (see **Table 1**). Of particular interest, RbClCu$_5$P$_2$O$_{10}$ shows double magnetic transitions with $T_{N1}$ =20 K and $T_{N2}$=7 K.[10] The structure of averievite contains kagome layers, which is a key feature of herbertsmithite ZnCu$_3$(OH)$_6$Cl$_2$,[13-16] making averievite very attractive for exploring quantum spin liquid via substitution or doping.[7] Density functional theory calculations predicted that CCCVO and CsClCu$_5$P$_2$O$_{10}$ can become quantum spin liquids when non-magnetic atoms such as Zn completely replace the Cu atoms in the triangle layers.[7, 12] Experimentally, long range magnetic order has been suppressed to below 2 K based on polycrystalline powders.[7, 17, 18]

Averievite compounds exhibit rich structure types and complex phase transitions.[9, 10] However, up to date, certain crystal structures and the underlying mechanism of phase transitions remain controversial. Theoretical calculations based on incorrect crystal structure could mislead experimenters. Furthermore, the existence of ambiguous crystal structures in the literature makes it difficult to understand the structure-property relationship. Six structure types (see **Table 1**) have been reported in averievite compounds: (i) $P3$ space group with $a_0=b_0\sim6.3$ Å and $c_0\sim8.4$ Å;[5] (ii) $P\bar{3}m1$ with $a=b=a_0$ and $c=c_0$;[6-10] (iii) $P\bar{3}$ with $a=b=2a_0$ and $c=c_0$;[9] (iv) $P3m$ with $a=b=\sqrt{3}a_0$, $c=c_0$;[6] (v) $P2_1/c$ with $a=c_0$, $b=a_0$, $c=\sqrt{3}a_0$ and $\beta\sim90°$;[7] and (vi) $C2/c$ with $a=\sqrt{3}a_0$, $b=b_0$, $c=2c_0$ and $\beta\sim95°$.[9, 10] Initially, the crystal structure of averievite at room temperature was solved and refined using $P3$ space group based on X-ray diffraction on natural single crystals.[5] In 2009, averievite compounds were first synthesized by Queen and the room temperature structure was reported to be $P\bar{3}m1$ for CCCVO and CsBrCu$_5$V$_2$O$_{10}$, and $P3m$ for RbClCu$_5$V$_2$O$_{10}$.[6] In 2018, CCCVO was reported to crystallize in the $P2_1/c$ space group at room temperature based on Rietveld refinements on high resolution synchrotron X-ray powder diffraction data and to undergo a phase transition from $P2_1/c$ to $P\bar{3}m1$ at ~310 K on warming.[7] In contrast, Kornyakov et al. reported $P\bar{3}$ space group at room tempeature for CCCVO with $a$ and $b$ doubling referring to the $P\bar{3}m1$ structure.[9] Besides CCCVO, the room temperature structure of (CuCl)$_2$Cu$_5$V$_2$O$_{10}$ is also under debate (**Table 1**), making it difficult to understand the structure-property relationship. On cooling, structural phase transitions have been reported in CCCVO ($P\bar{3}m1\rightarrow P2_1/c$),[7] CsBrCu$_5$P$_2$O$_{10}$ ($P\bar{3}m1\rightarrow P2_1/c$),[8] CsICu$_5$P$_2$O$_{10}$ ($P\bar{3}m1\rightarrow P2_1/c$)[8] and RbClCu$_5$P$_2$O$_{10}$ ($P\bar{3}m1\rightarrow C2/c$)[10] on the basis of powder diffraction. Due to the ambiguity of room temperature structure, the nature of the structural transitions in averievites remains elusive. Thus, high resolution synchrotron X-ray single crystal diffraction is needed to address the above-mentioned fundamental issue. Up to date, a number pf averievite compounds with dimensions of submillimeter on edge have been synthesized in the single crystal form, including CsClCu$_5$P$_2$O$_{10}$ (0.12×0.12×0.03 mm$^3$),[9] CCCVO (0.06×0.02×0.02 mm$^3$, 0.42×0.40×0.05 mm$^3$),[6, 9] RbClCu$_5$P$_2$O$_{10}$ (0.21×0.16×0.13 mm$^3$),[9] RbClCu$_5$V$_2$O$_{10}$ (0.11×0.04×0.04 mm$^3$),[6] KBrCu$_5$P$_2$O$_{10}$ (0.21×0.19×0.06 mm$^3$),[9] CsBrCu$_5$V$_2$O$_{10}$ (0.10×0.03×0.03 mm$^3$),[6] KClCu$_5$P$_2$O$_{10}$ (0.14×0.10×0.07 mm$^3$)[9] and (Cu$_{0.87}$Cl$_{1.035}$)$_2$Cu$_5$V$_2$O$_{10}$.[9] Bulk single crystals are highly demanded for investigation of direction dependent physical properties and for understanding the structure-property relationship.

**Table 1.** Summary of structural and magnetic properties of averievite. Note that s is for single crystal data, and p is for powder data.

| Averievites | Structure at 296 K | Phase Transition(s) T (K) | $T_N$ (K) | Ref. |
| --- | --- | --- | --- | --- |
| CCCVO | $P\bar{3}m1$ (s) | | - | [6] |
| | $P2_1/c$ (p) | 310, 125 ($P\bar{3}m1$ for T>310, $P2_1/c$ for 125<T<310) | 24 | [7] |
| | $P\bar{3}$ (s) | | - | [9] |
| CsClCu$_5$P$_2$O$_{10}$ | $P\bar{3}m1$ (p) | 12 ($P\bar{3}m1$ for T>12) | 3.8 | [8] |
| | $P\bar{3}m1$ (s) | - | - | [9] |
| KClCu$_5$P$_2$O$_{10}$ | $C2/c$ (s) | - | - | [9] |
| KBrCu$_5$P$_2$O$_{10}$ | $C2/c$ (s) | - | - | [9] |
| RbClCu$_5$P$_2$O$_{10}$ | $C2/c$ (s) | - | - | [9] |
| | $C2/c$ (p) | 310 ($P\bar{3}m1$ for T>310, $C2/c$ for T<310) | 20, 7 | [10] |
| RbClCu$_5$V$_2$O$_{10}$ | $P3m$ (s) | - | - | [6] |
| CsBrCu$_5$P$_2$O$_{10}$ | $P\bar{3}m1$ (p) | 75 ($P2_1/c$ for T<75) | 7 | [8] |
| CsBrCu$_5$V$_2$O$_{10}$ | $P\bar{3}m1$ (s) | - | - | [6] |
| CsICu$_5$P$_2$O$_{10}$ | $P\bar{3}m1$ (p) | 230 ($P2_1/c$ for T<230) | 12.5 | [8] |
| (CuCl)$_2$Cu$_5$V$_2$O$_{10}$ | $P\bar{3}m1$ (s) | - | - | [9] |
| | $P\bar{3}m1$ (p) | - | - | [10] |
| | P3 (s) | - | - | [5] |

In this contribution, we reported for the first time the growth of bulk single crystals of CCCVO using the flux method with CsCl/CuCl$_2$ as flux. Combined synchrotron X-ray single crystal and high-resolution powder diffraction, the room temperature structure of CCCVO was unambiguously determined to be $P\bar{3}$. A structural transition occurs at 305 K on cooling, resulting in a doubling of the cell parameters $a$ and $b$ and the crystal structure transforms from $P\bar{3}m1$ to $P\bar{3}$. Direction dependent magnetic susceptibility data were collected due to the availability of bulk single crystals. Thermal hysteresis in out-of-plane magnetic susceptibility around 305 K suggested a first-order structural transition and a cusp in both in-plane and out-of-plane magnetic susceptibility at 24 K indicates antiferromagnetic ordering. Two anomalies in heat capacity support structural transition at 127 K and antiferromagnetic transition at 24 K.

## 2. Experimental Section

**2.1 Polycrystalline preparation.** Polycrystalline samples of CCCVO were synthesized by solid-state reaction. CuO (Macklin, AR), V$_2$O$_5$ (Macklin, 99.99%) and CsCl (Macklin, AR) were weighed in the molar ratio of 5:1:1.01, mixed, ground, then pressed in pellets, and sintered at 500 °C for 12 h in air, followed by furnace cooling to room temperature. The pellets were ground, pressed, and sintered three times using the same conditions.

**2.2 Single crystal growth.** Single crystals of CCCVO were grown using the flux method in air-tight quartz tubes. Polycrystalline samples of CCCVO were mixed with flux consisting of CsCl/CuCl$_2$. The CuCl$_2$ was obtained by baking CuCl$_2$·2H$_2$O (Aladdin, AR) at 150 °C for 12 h. The mixture was sealed in quartz tubes, heated to 600 °C, held for 1 day, and then cooled to 400 °C at a rate of 2 °C/h, followed by furnace cooling to room temperature. Finally, black hexagonal single crystals were obtained by removing excess flux using deionized water. **Table 2** lists the growth conditions of CCCVO single crystals.

**Table 2.** Growth conditions of CCCVO crystals.

| No. | Solute | Solvent CsCl/CuCl$_2$ (molar ratio) | Solute: Solvent | Dwelling time (h) at 600 °C | Cooling range (°C) and time (h) | Result |
|---|---|---|---|---|---|---|
| | | | | | | |

| | | (weight ratio) | | | | |
|---|---|---|---|---|---|---|
| 1 | CuO | 5:5 | 10:1 | 24 | 600-400, 48 | CCCVO single crystals (1.2 mm on edge), $Cu_2V_2O_7$ |
| 2 | CCCVO | 5:5 | 5:1 | 24 | 600-400, 48 | CCCVO single crystals (1.2 mm on edge), $Cu_2V_2O_7$ |
| 3 | CCCVO | 4:6 | 5:1 | 24 | 600-400, 48 | CCCVO single crystals (1.1 mm on edge), $Cu_2V_2O_7$ |
| 4 | CCCVO | 6:4 | 5:1 | 24 | 600-400, 48 | CCCVO single crystals (1.2 mm on edge), $Cu_2V_2O_7$ |
| 5 | CCCVO | 7:3 | 5:1 | 24 | 600-400, 48 | CCCVO single crystals (1.4 mm on edge), $Cu_2V_2O_7$ |
| 6 | CCCVO | 8:2 | 5:1 | 24 | 600-400, 48 | CCCVO single crystals (0.9 mm on edge), $Cu_2V_2O_7$ |
| 7 | CCCVO | 9:1 | 5:1 | 24 | 600-400, 48 | CCCVO single crystals (1.5 mm on edge), $Cu_2V_2O_7$ |
| 8 | CCCVO | 7:3 | 5:1 | 24 | 600-400, 96 | CCCVO single crystals (2.4 mm on edge), $Cu_2V_2O_7$ |
| 9 | CCCVO | 7:3 | 5:1 | 24 | 600-400, 168 | CCCVO single crystals (3.4 mm on edge), $Cu_2V_2O_7$ |

**2.3 In-house X-ray powder diffraction (PXRD).** Bruker AXS D2 Phaser X-ray powder diffractometer with Cu-$K_\alpha$ radiation ($\lambda$ = 1.5418 Å) was used to check phase purity. Data of solid-state reaction were collected in the $2\theta$ range of 5-70° with a step size of 0.02° and a step time of 0.1 s. Data for Rietveld refinements were collected on pulverized single crystals of CCCVO in the $2\theta$ range of 5-140° with a step size of 0.01° and a time of 1.0 s. TOPAS V6 was used to perform Rietveld refinements. Refined parameters include background (Chebyshev function, 10 order), sample displacement, lattice parameters, crystallite size L, and strain G.

**2.4 Synchrotron X-ray single crystal diffraction (SXRD).** Single crystals were mounted on the tips of glass fibers, and measurements were taken using a Huber 3-circle diffractometer. The diffraction data were collected using a Pilatus (CdTe) 1M area detector with synchrotron radiation ($\lambda$=0.41328 Å) at temperatures of 400 K, 350 K, 296 K, and 200 K, as well as between 300 K and 105 K in 5 K intervals during cooling. These measurements were conducted at Beamline 15-ID-D (NSF's ChemMatCARS) at the Advanced Photon Source, Argonne National Laboratory. Bruker APEX4 software was used for indexing, data reduction and image processing.[19] The structure was solved by direct methods and refined with full matrix least-squares methods on $F^2$. The atoms were modelled using anisotropic ADPs, and the refinements converged for I>2σ(I), where I is the intensity of the reflections and σ(I) is the standard deviation. The calculations were carried out using the SHELXTL crystallographic software package.[20] **Table 3** summarizes the crystal parameters, data collection and details of structure refinement at 400, 350, 296 and 200 K, and **Table S1** lists selected bond lengths (Å) and angles (°). Further details of the crystal structure investigations may be obtained from the joint CCDC/FIZ Karlsruhe online deposition service by quoting the deposition numbers CSD 2338862, 2264551, 2264817, and 2264549.

**Table 3.** Crystallographic data and refinement parameters for CCCVO at 200 K, 296 K, 350 K and 400 K.

| Empirical formula | CsClCu$_5$V$_2$O$_{10}$ | | | |
|---|---|---|---|---|
| Formula weight (g/mol) | 747.94 | | | |
| Temperature (K) | 200(2) | 296(2) | 350(2) | 400(2) |
| Wavelength (Å) | 0.41328 | | | |
| Crystal system | Trigonal | | | |
| Space group | $P\bar{3}$ | $P\bar{3}$m1 | $P\bar{3}$m1 | $P\bar{3}$m1 |
| Unit cell parameters | $a$ = 12.5749(3) Å<br>$b$ = 12.5749(3) Å<br>$c$ = 8.2804(3) Å | $a$ = 6.29890(10) Å<br>$b$ = 6.29890(10) Å<br>$c$ = 8.2854(3) Å | $a$ = 6.30110(10) Å<br>$b$ = 6.30110(10) Å<br>$c$ = 8.2859(2) Å | $a$ = 6.3050(3) Å<br>$b$ = 6.3050(3) Å<br>$c$ = 8.2872(5) Å |
| Volume (Å$^3$) | 1133.94(7) | 284.691(14) | 284.907(11) | 285.30(3) |
| Z | 4 | 1 | 1 | 1 |
| Density (calculated) (g/cm$^3$) | 4.381 | 4.363 | 4.359 | 4.353 |
| Absorption coefficient (mm$^{-1}$) | 3.190 | 3.177 | 3.174 | 3.170 |
| F (000) | 1372.0 | 343.0 | 343.0 | 343.0 |
| Crystal size (mm$^3$) | 0.025 × 0.025 × 0.015 | | | |
| Theta range for data collection (°) | 1.087 to 18.521 | 1.429 to 22.474 | 1.429 to 22.052 | 1.429 to 20.135 |
| Index ranges | -19≤$h$≤19,<br>-19≤$k$≤19, -11≤$l$≤12 | -9≤$h$≤11,<br>-11≤$k$≤9, -15≤$l$≤11 | -10≤$h$≤9,<br>-11≤$k$≤9, -15≤ $l$≤11 | -9≤$h$≤10,<br>-10≤$k$≤9, -11≤1 ≤13 |
| Reflections collected | 41794 | 13317 | 12977 | 8179 |
| Independent reflections | 2887<br>[R(int) = 4.91%] | 767<br>[R(int) = 4.3%] | 721<br>[R$_{int}$ = 4.88%] | 564<br>[R$_{int}$ = 5.55%] |
| Completeness to theta = 14.357° | 100.00% | 100.00% | 100.00% | 100.00% |
| Absorption correction | multi-scan | | | |
| Max. and min. transmission | 0.827 and 0.934 | 0.854 and 0.942 | 0.857 and 0.942 | 0.796 and 0.907 |
| Refinement method | Full-matrix least-squares on F$^2$ | | | |
| Data / restraints / parameters | 2887/0/135 | 767 / 0 / 31 | 721/0/32 | 564/0/31 |
| Goodness-of-fit on $F^2$ | 1.001 | 1.184 | 1.161 | 1.186 |
| Final R indices [I>2σ(I)] | $R_1$= 8.15 %, $wR_2$= 20.97 % | $R_1$= 2.68 %, $wR_2$= 7.38 % | $R_1$ = 2.83 %, $wR_2$ = 7.54 % | $R_1$ = 4.31 %, $wR_2$ = 12.77 % |
| R indices (all data) | $R_1$= 8.84 %, $wR_2$= 21.70 % | $R_1$= 2.84 %, $wR_2$= 7.45 % | $R_1$ = 3.02 %, $wR_2$ = 7.66 % | $R_1$ = 4.44 %, $wR_2$ = 12.98 % |
| Largest diff. peak and hole (e·Å$^{-3}$) | 6.31 and -1.75 | 2.72 and -2.13 | 2.66 and -1.98 | 1.71 and -2.49 |
| CSD number | 2338862 | 2264551 | 2264817 | 2264549 |

**2.5 High-resolution synchrotron X-ray powder diffraction (HRPXRD).** HRPXRD data for CCCVO were collected in the 2θ range of 0.5-50° with a step size of 0.001° and a step time of 0.1 s with X-ray wavelength of λ = 0.45903 Å at temperatures of 100 K, 200 K, 295 K and 400 K at Beamline 11-BM at the Advanced Photon Source, Argonne National Laboratory. Well-ground powders from crushing single crystals were loaded into a Kapton capillary tube with a diameter of 0.8 mm. The tube was then mounted on a magnetic sample base, which was used in the beamline sample changer. During data acquisition, the sample was rotated continuously at 5600 rpm. Temperature was controlled using an Oxford Cryostream 700 Plus $N_2$ gas blower. Data were first recorded at 295 K, and then the sample was cooled to 100 K and data were collected on warming (100, 200, and 400 K). The data were analyzed using the Rietveld method with GSAS II software.[21] Crystal structures obtained from SXRD were used as starting models, and the refined parameters include scale, background, unit cell parameters, domain size, microstrain, atomic positions and thermal parameters. Isotropic domain size and generalized microstrain models were used.

**2.6 Magnetic Susceptibility.** The Quantum Design MPMS3 was used to collect DC susceptibility data with a single crystal of 8.8 mg. ZFC-W (zero-field cooling, data collection on warming), FC-C (field cooling, data collection on cooling) and FC-W (field cooling, data collection on warming) were collected between 1.8 and 350 K with a heating rate of 3 K/min under an external magnetic field of 0.2 T with the magnetic field parallel/perpendicular to the *ab* plane. ZFC, FC-C and FC-W were collected between 1.8 and 40 K with a heating rate of 3 K/min under an external magnetic field of 0.01, 0.5, 1.0, 2.0, 3.0, 4.0, 5.0, 6.0, and 7.0 T with the magnetic field parallel/perpendicular to the *ab* plane. Magnetization data as a function of the field were collected at intervals of 0.1 T between -7 T to +7 T at 1.8, 2.0, 5.0, 10.0, 20.0, 30.0 and 300 K with magnetic field parallel/perpendicular to the *ab* plane. Multiple single crystals were measured.

**2.7 Heat capacity ($C_P$).** The heat capacity of a CCCVO single crystal was measured in a Quantum Design PPMS using the relaxation method in the range of 2 K - 200 K. Apiezon-N vacuum grease was used to secure the crystal (6.7 mg) to the sapphire sample stage. Data in the 18 K - 28 K range were also collected under magnetic fields of 3 T and 7 T. The specific heat contribution from the sample holder platform and grease was determined before mounting the sample and subtracted from the total heat capacity.

## 3. Results and Discussion

**3.1 Flux growth of bulk single crystals.** Polycrystalline powders of CCCVO were synthesized using solid-state reactions (see **Figure S1**) according to the recipe reported by A. S. Botana.[7] In order to check whether CCCVO is congruent or incongruent melting, as-synthesized polycrystalline powders were heated to 800 °C in air, and the residual materials after heating were checked using X-ray powder diffraction. It was found that no CCCVO was left and most peaks were indexed to CuO (see **Figure S1**), indicating that CCCVO is incongruent melting. Thus, melt growth techniques including floating zone, Bridgman and Czochralski are impossible, and flux method was selected for single crystal growth.

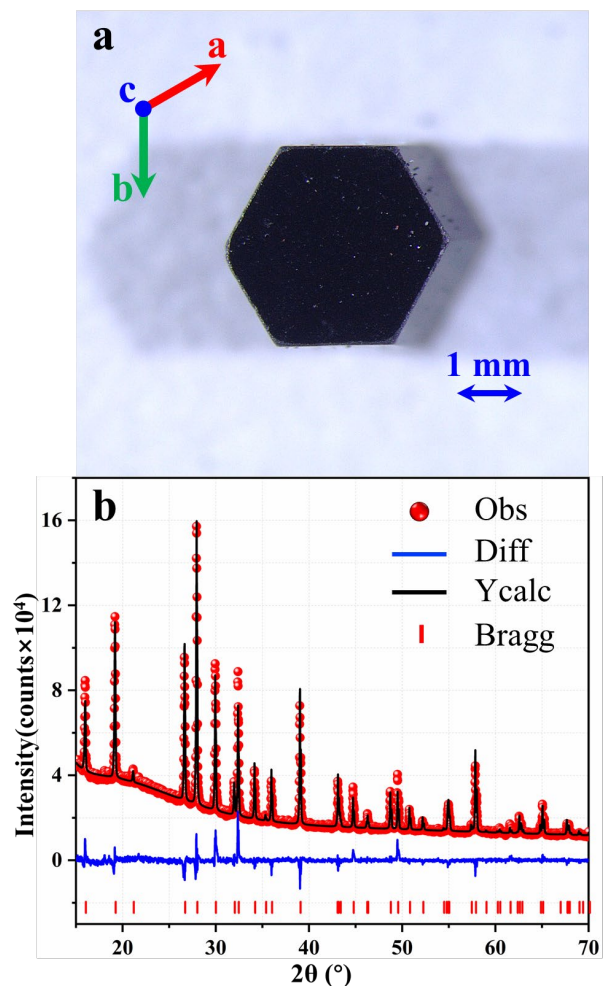

**Figure 1.** (a) Photo of a typical single crystal of CCCVO from flux growth (3.37×2.93×0.92 mm$^3$). (b) In-house X-ray powder diffraction pattern of pulverized single crystals of CCCVO and Rietveld refinement using $P\bar{3}m1$.

The first and most difficult step for flux growth is finding a suitable flux.[22-25] Initially, we attempted to grow single crystals of CCCVO using the conditions reported by Queen et al;[6] however, we failed. We then turned to self-fluxes. The melting points of CsCl and CuCl$_2$ are 645 and 620 °C, respectively. Attempting to grow CCCVO using these two materials separately did not result in CCCVO. We then explored different combinations of CsCl and CuCl$_2$, and found that their combination in the molar ratio of 1:1 has a low eutectic point of about 420 °C. CuO was added to the CsCl/CuCl$_2$ = 5:5 flux system according to the weight ratio of 1:5, and the cooling procedure was 600-400 °C for 48 h. After the excess flux was dissolved in water, hexagonal and elongated black crystals were obtained (see **Figure S2**). **Figure 1** shows the in-house powder X-ray diffraction data of pulverized hexagonal single crystals. All peaks can be indexed to space group $P\bar{3}m1$, and Rietveld refinement using our single crystal model converged to $R_{wp}$ = 4.24%, $R_p$ = 2.9%, GOF = 2.08 with lattice parameters of $a = b$ = 6.36402(9) Å and $c$ = 8.37352(15) Å. Our result is consistent with the report by Queen et al.[6]

In order to obtain large size and high-quality single crystals, we optimized the growth conditions including the raw material, molar ratio of CsCl to $CuCl_2$, and cooling time. The growth conditions and corresponding results are listed in **Table 2**. Firstly, we replaced CuO with CCCVO polycrystalline powders. Polycrystalline powders are frequently employed as precursors for the growth of single crystals of complex compounds by the flux method.[26, 27] There is no change in the dimensions of the as-grown single crystals; however, the amount of secondary phases decreased significantly. Subsequently, in order to further optimize the growth conditions, the CsCl: $CuCl_2$ ratio was adjusted. As the concentration of CsCl increases, the amount of the second phase diminishes. At CsCl: $CuCl_2$=9:1, the crystal size is up to 1.5 mm, but minor impurity adheres to the crystal surface and is difficult to remove. CsCl: $CuCl_2$=7:3 is the best ratio for growing crystals. Finally, in order to obtain larger size single crystals, the cooling time from 600 to 400 °C was optimized. By extending cooling time to 96 h, crystal size increased up to 2.4 mm on edge, and further to 3.4 mm when prolonged to 168 h. **Figure 1 inset** shows a photograph of a typical CCCVO crystal with crystallographic directions labeled.

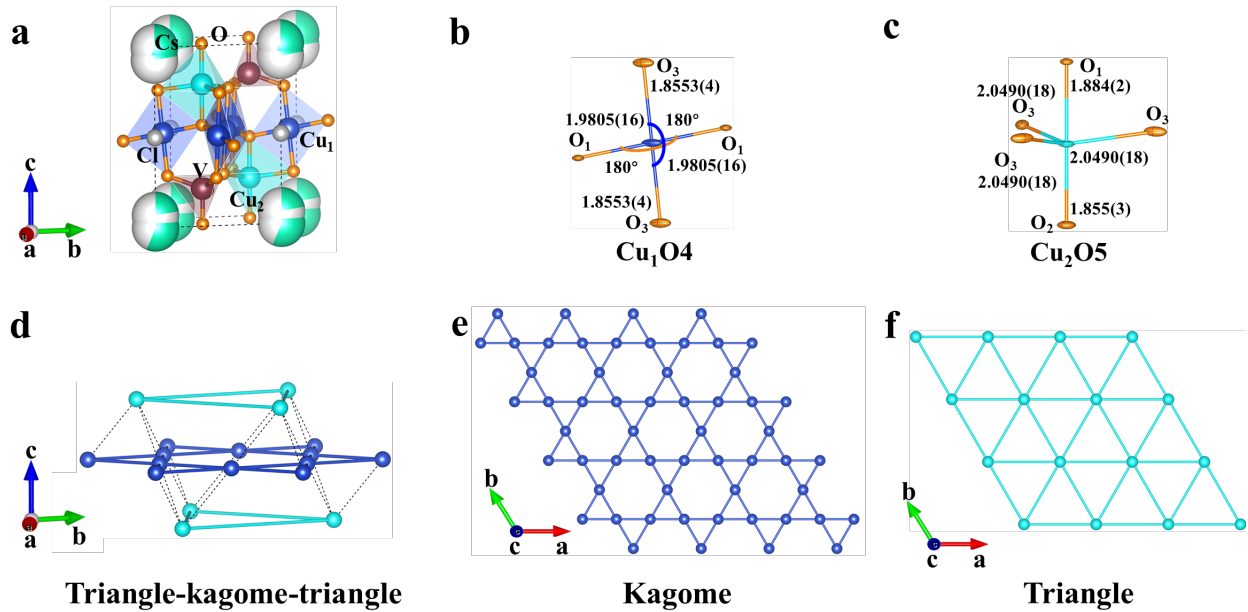

**Figure 2.** Crystal structure of CCCVO at 350 K. (a) 3D crystal structure viewed along *a* axis. (b) Ball-and-stick drawing of copper oxygen planar environment. (c) Ball-and-stick drawing of copper oxygen polyhedral. (d) Triangle-kagome-triangle trilayer consisting of copper atoms. (e) Kagome layer consists of Cu1 atoms. (f) Triangle layer consists of Cu2 atoms.

### 3.2 Structural transition at ~ 305 K.

**3.2.1 Single crystal X-ray diffraction at various temperatures.** Single crystal XRD data were collected above and below 305 K to investigate the phase transition previously believed as $P\bar{3}m1$-$P2_1/c$ transition.[7] The structure at 400 and 350 K were solved using $P\bar{3}m1$. As we will discussed later, the superlattice peaks at 296 K were too weak so we solved the structure using $P\bar{3}m1$. **Figure**

**2a** shows the ball-and-stick model at 350 K. CCCVO belongs to the trigonal space group $P\bar{3}m1$ with cell parameters of $a = b = 6.30110(10)$ Å, $c = 8.2859(2)$ Å, and $Z = 1$. The asymmetric unit contains two Cs, one V, two Cu, one Cl, and three oxygen atoms. There are two crystallographic sites for Cu, the first site (Cu1) is bonded to four oxygen atoms with bond lengths of Cu1-O1 = 1.8553(4) Å and Cu1-O3 = 1.9805(16) Å forming a planar environment (see **Figure 2b**). The copper atom (Cu2) at the second site is coordinated by five oxygen atoms with bond lengths of 1.855(3) Å, 1.884(2) Å and 2.0490(18) Å (see **Figure 2c**). V is surrounded by four oxygens with a bond length of 1.644(3)-1.7054(18) Å and forms a tetrahedral geometry. **Figure 2d-f** illustrates the Cu lattice at 350 K (See **Table S3** for details). Cu1 atoms form a Kagome layer with a neighboring Cu1-Cu1 distance of 3.15055(6) Å, while Cu2 atoms form two triangular layers with 6.30110(11) Å between adjacent Cu2. The kagome layer is sandwiched by the two triangular layers with Cu1-Cu2 distance of 2.8928(4) Å, forming a triangle-kagome-triangle trilayer. Cs and Cl atoms are filled among trilayers to balance the charge.

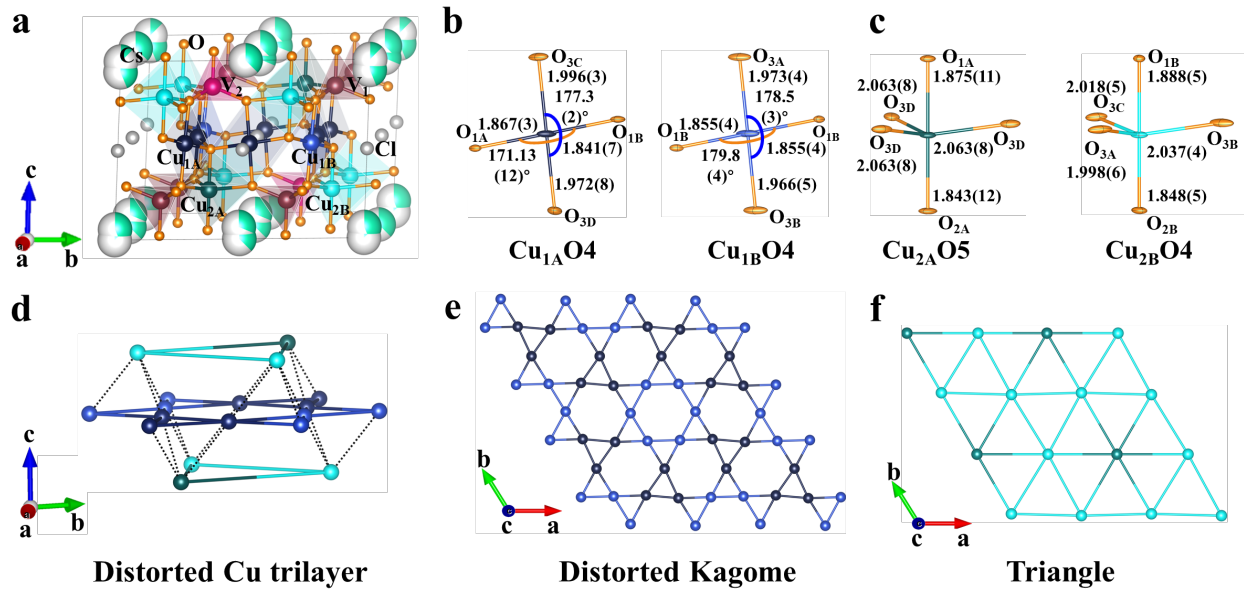

**Figure 3.** Crystal structure of CCCVO at 200 K. (a) 3D crystal structure viewed along $a$ axis. (b) Ball-and-stick drawings of copper oxygen tetrahedra. (c) Ball-and-stick drawings of copper oxygen polyhedral. (d) Triangle-kagome-triangle trilayer consisting of copper atoms. (e) Kagome layer consisting of Cu1. (f) Triangle layer consisting of Cu2.

**Figure 3** presents the crystal structure of CCCVO at 200 K. It belongs to the $P\bar{3}$ space group of the trigonal crystal system with cell parameters of $a = b = 12.5749(3)$ Å, $c = 8.2804(3)$ Å, and $Z = 4$. Compared with the data at 350 K, apparent superlattice peaks with $\mathbf{Q} = (1/2, 1/2, 0)$ based on unit cell setting with $a=b\sim6.28$ Å and $c\sim8.28$ Å were observed at 200 K (see **Figure 4**). These superlattice peaks cannot be indexed using the monoclinic cell ($P2_1/c$ with $a\sim8.37$ Å, $b\sim6.36$ Å, $c\sim11.01$ Å and $\beta\sim90.02°$) reported by Botana et al.,[7] suggesting a different structure than $P2_1/c$. The O3 site at high temperature splits into four crystallographic sites, and the sites occupied by all other atoms split into two sites, thus the asymmetric unit at 200 K consists of four Cs atoms, four Cu atoms, two V atoms, two Cl atoms, and eight O atoms. **Figure 3b** shows the local environments of Cu-O polyhedral with bond distance and angles. Notably, the two-dimensional Cu1-O planar environment at high temperature distorts and becomes three-dimensional. Cu2 splits into Cu2A and Cu2B (**Figure 3c**). V atoms are bonded to four oxygen atoms with the bond lengths ranging from 1.650(5) to 1.727(5) Å. Cs and Cl ions fill the voids to realize the charge balance in the system. **Figure 3d** illustrates the distorted triangle-kagome-triangle trilayer consisting of Cu at 200 K (See **Table S3** for details). The distorted kagome layer as shown in **Figure 3e** consists of Cu1A and Cu1B. Within the kagome layer, the three neighboring Cu1A form an equilateral triangle with Cu1A-Cu1A distance of 3.171(3) Å, and six neighboring Cu1B form hexatomic rings with Cu1B-Cu1B distance of 3.1551(14) Å. In addition, the nearest distance between Cu1A and Cu1B are 3.124(3) Å and 3.150(3) Å, the interior angles of triangle Cu1A-Cu1B-Cu1B are 59.41(7)°, 60.22(8)°, and 60.37(6)°, and the interior angles of a hexagon formed by four Cu1A and two Cu1B are 111.19(8)° × 2, 120.37(6)° × 2, and 128.44(10)° × 2, deviating strongly from 120°. **Figure 3f** shows the triangular layer formed by Cu2A and Cu2B. Along $a$ direction, there are two arrangements of Cu2: first, Cu2A alternates with Cu2B, and second, all occupied by Cu2A. Neighboring Cu2A-Cu2A has a distance of 6.284(3) Å, and Cu2A-Cu2B of 6.1708(15) Å and 6.4041(15) Å. Our structural model is consistent with Kornyakov et al. at 296 K.[9]

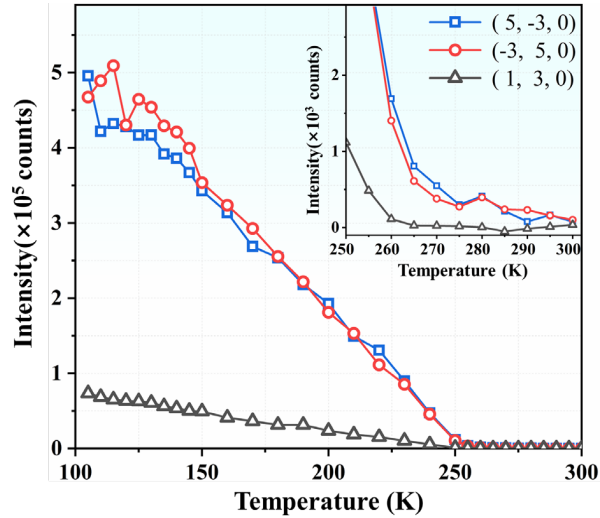

**Figure 4.** The integrated intensity of (5, -3, 0), (-3, 5, 0) and (1, 3, 0) from synchrotron X-ray single crystal diffraction based on $P\bar{3}$ as a function of temperature from 300 K to 105 K on cooling. Inset: Zoom-in between 250 K and 300 K.

**3.2.2. Variable-temperature single-crystal X-ray diffraction.** Our single crystal data show different structures across 350 K. One natural question is what temperature the phase transition occurs. To address this question, we collected single-crystal diffraction data on cooling from 300 to 105 K. **Figure 4** show the temperature dependence of three typical peaks indexed and integrated using the 200 K unit cell as a starting point. Two anomalies are observed, the first one at ~ 270 K and the second one at ~ 127 K, indicating that CCCVO single crystals undergo two successive structural transitions. The transition at ~270 K is lower than 305 K, probably due to the small counting time during data collection. If we increase statistics, the transition temperature is expected to approach 305 K. By refining data collected at various temperatures, we obtained cell parameters $a$, $c$, and volume as a function of temperature, as shown in **Figure S3**.

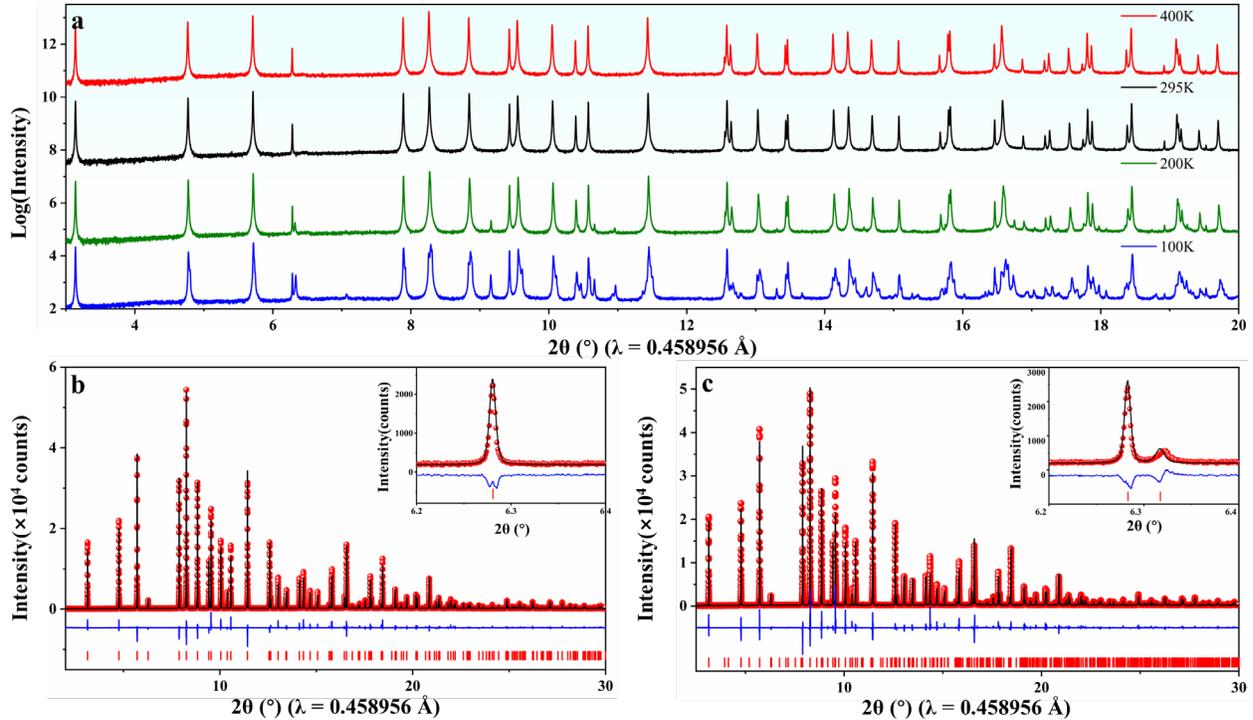

**Figure 5.** Variable temperature synchrotron X-ray powder diffraction patterns of CCCVO and Rietveld refinement. (a) High-resolution synchrotron X-ray powder diffraction pattern of CCCVO in the 2$\theta$ range of 3-20° measured at 100, 200, 295 and 400 K. (b) Rietveld refinement on the high-resolution synchrotron X-ray powder diffraction data of CCCVO collected at 400 K in the 2θ range of 2-30° using $P\bar{3}m1$. (c) Rietveld refinement on the high-resolution synchrotron X-ray powder diffraction data of CCCVO collected at 200 K in the 2θ range of 2-30° using $P\bar{3}$.

**3.2.3. HRPXRD and Rietveld refinement at various temperatures.** To verify our single crystal model, we conducted temperature dependent high resolution powder X-ray diffraction at 11-BM at APS. **Figure 5a** presents the high-resolution synchrotron X-ray diffraction data at 400, 295, 200 and 100 K. Clearly, the patterns at 400 and 295 K are identical. In contrast, additional peaks appear at 200 K compared with of 295 K, indicating structural transition above 200 K. Upon further cooling, more extra peaks show up at 100 K, suggesting another structural phase transition between 200 and 100 K. These high-resolution powder X-ray diffraction data are consistent with our single crystal results discussed previously. We then carried out Rietveld refinements on the 400 and 200 K data using single crystal structural models (**Table S2**). **Figure 5b** shows the structural refinement of 400 K data utilizing $P\bar{3}m1$. The refinement converged to $R_{wp}$ = 9.678% and GOF = 1.87 with lattice parameters of $a = b$ = 6.371681(8) Å and $c$ = 8.378223(6) Å. **Figure 5c** presents the refinement of 200 K data utilizing $P\bar{3}$, and the refinement converged to $R_{wp}$ = 13.447% and GOF = 2.82 with lattice parameters of $a = b$ = 12.72509(4) Å and $c$ = 8.375747(12) Å. The excellent agreement between calculated and observed intensity at 400 K and 200 K corroborates our single crystal X-ray diffraction results.

**3.2.4 Magnetic susceptibility as a function of temperature. Figure 6** presents the in-plane and out-of-plane magnetic susceptibility data of CCCVO single crystals as a function of temperature. **Figure 6a inset** shows a thermal hysteresis in the range of 290 - 310 K between FC-C and FC-W, suggesting a first-order phase transition. This is consistent with our synchrotron X-ray single crystal diffraction result.

**3.3 Structural transition at ~ 127 K.** Heat capacity data show a clear anomaly at 127 K (**Figure 7a**). Combining X-ray single crystal diffraction, powder diffraction and heat capacity, a structural transition across 127 K occurs. However, we haven't obtained a satisfied structural model for the structure below this transition due to poor data quality at 100 K (e.g., twinning and cracking). We tried our best to refine our high-resolution powder diffraction data using structural models reported for averievites in the literature.[7,9,10] We first tried to refine using a single phase. **Figures S4 to S6** show the Le-Bail fit using $P\bar{3}$, $P2_1/c$, and $C2/c$ as the initial model, and the refinements converged to $R_{wp}$ = 29.73%, 28.33%, and 29.66%, respectively. As can be seen, none of them is satisfactory. Because 100 K is very close to the transition temperature, it is likely that the 100 K data consists of multiple phases. Thus, $P\bar{3}$ and $P2_1/c$, $P\bar{3}$ and $C2/c$ were used as initial models (**Figure S7 and S8**), and the fit converged to $R_{wp}$ = 30.22% and 27.72%, respectively. Therefore, synchrotron X-ray single crystal diffraction at a low temperature of less than 100 K is needed to address this puzzle.

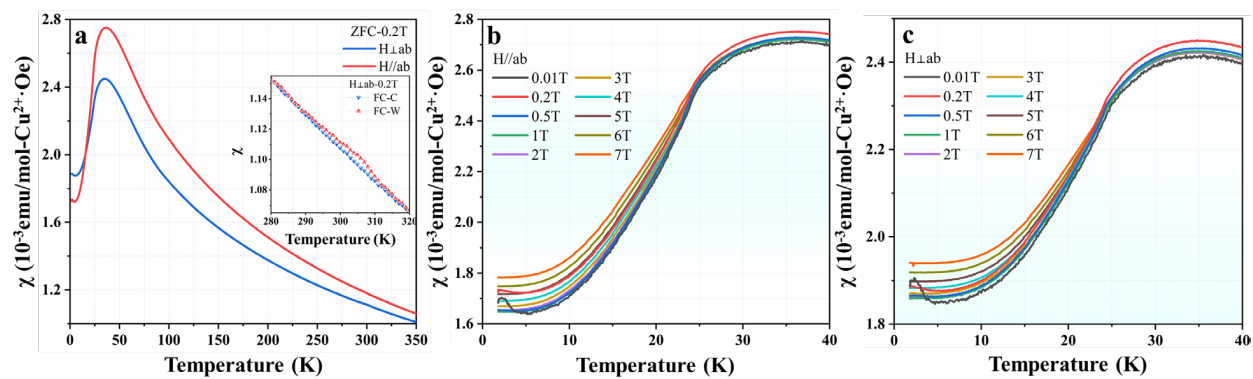

**Figure 6.** (a) Magnetic susceptibility (ZFC) as a function of temperature with magnetic field of 0.2 T parallel to the *ab* plane and perpendicular to the *ab* plane. Inset: Magnetic susceptibility as a function of temperature with magnetic field of 0.2 T perpendicular to the *ab* plane in the range of 280-320 K. (c) Magnetic susceptibility (ZFC) as a function of temperature with magnetic field parallel to the *ab* plane. (d) Magnetic susceptibility (ZFC) as a function of temperature with magnetic field perpendicular to the *ab* plane.

**3.4 Antiferromagnetic transition at ~24 K.**

**3.4.1 MT and field-dependence.** The availability of bulk single crystals makes it possible to investigate direction dependent physical properties. **Figure 6** shows the magnetic susceptibility of CCCVO as a function of temperature with an external magnetic field of 0.2 T parallel/perpendicular to the *ab* plane. The two curves are not overlapping, indicating anisotropic magnetic properties. The first derivatives of ZFC **(Figure S9)** show a maximum at 24 K, indicating a phase transition. The transition temperature is consistent with Botana et al.[7] **Figures 6b, c** present the temperature-dependent magnetic susceptibility (ZFC) at various magnetic fields in the temperature range of 1.8 K - 40 K. The cusp shape at 24 K and obvious response to the external magnetic field suggest an antiferromagnetic transition. **Figure S10** presents the Curie-Weiss fit in the temperature range between 150 K and 275 K with H//ab, an effective moment of $\mu_{eff}$=2.09 $\mu_B$ and Weiss temperature of $\theta_{CW}$= -161 K are obtained. **Figure S11** presents the Curie-Weiss fit with H⊥ab, an effective moment of $\mu_{eff}$= 2.13 $\mu_B$ and Weiss temperature of $\theta_{CW}$= -213 K are obtained.

The effective magnetic moments are close to the expected value of 1.73 $\mu_B$ for the Cu ions with S = 1/2,[28] and the negative and large values of the Weiss temperatures indicate a strong antiferromagnetic interaction in the system. Magnetization data as a function of magnetic fields are shown in **Figures S12 and S13**.

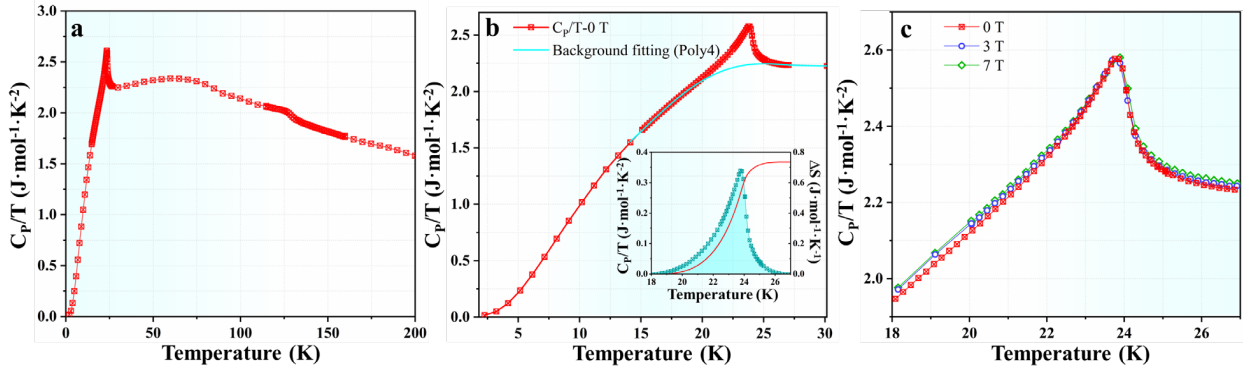

**Figure 7.** Heat capacity of CCCVO at various fields. (a) Heat capacity of CCCVO in the range of 2-200 K at 0 T. (b) The specific heat exhibits a variation with temperature within the range of 2 to 30 K at 0 T (red square dot line). The cyan line represents the background fit of the fourth-order polynomial. Inset: cyan point line plot represents the discrepancy between the actual and the fitted background, the red line represents the integration. (c) The specific heat as a function of magnetic field.

**3.4.2 Heat capacity. Figure 7** shows the specific heat of a CCCVO single crystal at various magnetic fields. The λ-type peak at 24 K (**Figure 7b**), which is consistent with the 1st-order derivative of the magnetic susceptibility, suggests a second-order phase transition. The specific heat background was fitted with a fourth-order polynomial, and the specific heat difference and its integral are shown in the inset of **Figure 7b**. The entropy change associated with the magnetic order is estimated to be ΔS=0.78 (J/mol·K), which is only 2.7% of the expected three-dimensional antiferromagnetic transition of 5Rln2. The heat capacity data under different applied magnetic fields are shown in **Figure 7c**, where the peaks overlap, indicating that a larger magnetic field is needed to suppress the transition.

## 4. Conclusion

We report the successful growth of bulk single crystals of CCCVO using the flux method. The structure at room temperature was determined to be $P\bar{3}m1$. Combining variable temperature synchrotron X-ray single crystal diffraction and high-resolution powder diffraction, two first-order structural transitions (one at ~305 K and another at ~127 K) were observed. The structure between 305 and 200 K was unambiguously solved using $P\bar{3}$ by synchrotron X-ray single-crystal diffraction. The triangle-kagome-triangle trilayer structure, composed of $Cu^{2+}$, is strongly distorted on cooling. Our experiments have resolved the long standing controversy of the crystal structure of averievite at room temperature and the nature of phase transition across ~305 K. Direction dependent magnetic susceptibility measurements show two anomalies, one between 290 and 310 K corresponding to a structural transition and another one at 24 K which corresponds to antiferromagnetic ordering. Heat capacity shows two anomalies: one at 127 K and the other one at 24 K. The preparation of large CCCVO crystals serves as a template for growing other averievite single crystals, including parent phases such as $CsClCu_5P_2O_{10}$, and quantum spin liquids candidates such as $CsClCu_3Zn_2V_2O_{10}$ and $CsClCu_3Zn_2Ti_2O_{10}$.[7] Our results not only reveal the nature of structural transitions and anisotropic magnetic properties in the averievite $CsClCu_5V_2O_{10}$ but also calls for more systematic study in this class of materials.

## ASSOCIATED CONTENT

**Supporting Information**.

The Supporting Information is available free of charge at ####. Table S1. Selected bond length and angles for $CsClCu_5V_2O_{10}$ at 200 K and 350 K. Table S2. Crystallographic data and Rietveld refinement on the high-resolution synchrotron X-ray powder diffraction data (11-BM, APS). Table S3. The bond lengths and angles of Cu lattice at 200 K and 350 K. Figure S1. In-house X-ray diffraction patterns of $CsClCu_5V_2O_{10}$ polycrystalline powders from solid-state reaction (Before) and resultant sample after treatment at 800 °C (After). Figure S2. X-ray diffraction pattern of pulverized hexagonal black crystals, and pulverized elongated black crystals. Figure S3. Lattice parameter as a function of temperature. Figure S4. Le Bail fit by TOPAS V6 on the high-resolution synchrotron X-ray powder diffraction data of CCCVO collected at 100 K in the 2θ range of 2-20° using $P\bar{3}$. Figure S5. Le Bail fit by TOPAS V6 on the high-resolution synchrotron X-ray powder diffraction data of CCCVO collected at 100 K in the 2θ range of 2-20° using $P2_1/c$. Figure S6. Le Bail fit by TOPAS V6 on the high-resolution synchrotron X-ray powder diffraction data of CCCVO collected at 100 K in the 2θ range of 2-20° using $C2/c$. Figure S7. Le Bail fit by TOPAS V6 on the high-resolution synchrotron X-ray powder diffraction data of CCCVO collected at 100 K in the 2θ range of 2-20° using $P\bar{3}$ and $P2_1/c$. Figure S8. Le Bail fit by TOPAS V6 on the high-resolution synchrotron X-ray powder diffraction data of CCCVO collected at 100 K in the 2θ range of 2-20° using $P\bar{3}$ and $C2/c$. Figure S9. The first derivatives of magnetic susceptibility (ZFC) as a function of temperature with a magnetic field parallel to the *ab* plane. Figure S10. Inverse susceptibility (1/χ) with magnetic field of 0.2 T parallel to the *ab* plane and the CW fit for data between 150 K and 275 K. Figure S11. Inverse susceptibility (1/χ) with magnetic field of 0.2 T perpendicular to the *ab* plane and the CW fit for data between 150 K and 275 K. Figure S12. Magnetization as a function of magnetic fields at various temperatures (1.8, 2, 3, 5, 10, 20, 30 and 300 K) with applied magnetic fields parallel to the *ab* plane. Figure S13. Magnetization as a function of magnetic fields at various temperatures (1.8, 2, 3, 5, 10, 20, 30 and 300 K) with applied magnetic fields perpendicular to the *ab* plane.

CCDC 2338862, 2264551, 2264817, and 2264549 contain the supplementary crystallographic data for this paper. These data can be obtained free of charge via http://www.ccdc.cam.ac.uk/data_request/cif, or by emailing data_request@ccdc.cam.ac.uk, or by contacting The Cambridge Crystallographic Data Centre, 12 Union Road, Cambridge CB2 1EZ, UK; fax: +44 1223 336033.

## AUTHOR INFORMATION

**Corresponding Author**

**Junjie Zhang** - State Key Laboratory of Crystal Materials and Institute of Crystal Materials, Shandong University, Jinan 250100, Shandong, China; E-mail: junjie@sdu.edu.cn

**Authors**

**Chao Liu** - State Key Laboratory of Crystal Materials and Institute of Crystal Materials, Shandong University, Jinan 250100, Shandong, China;

**Chao Ma** - State Key Laboratory of Crystal Materials and Institute of Crystal Materials, Shandong University, Jinan 250100, Shandong, China;

**Tieyan Chang** - NSF's ChemMatCARS, The University of Chicago, Lemont, IL 60439, United States

**Xiaoli Wang** - State Key Laboratory of Crystal Materials and Institute of Crystal Materials, Shandong University, Jinan 250100, Shandong, China;

**Chuanyan Fan** - State Key Laboratory of Crystal Materials and Institute of Crystal Materials, Shandong University, Jinan 250100, Shandong, China;

**Lu Han** - State Key Laboratory of Crystal Materials and Institute of Crystal Materials, Shandong University, Jinan 250100, Shandong, China;

**Feiyu Li** - State Key Laboratory of Crystal Materials and Institute of Crystal Materials, Shandong University, Jinan 250100, Shandong, China;

**Shanpeng Wang** - State Key Laboratory of Crystal Materials and Institute of Crystal Materials, Shandong University, Jinan 250100, Shandong, China;

**Yu-Sheng Chen** - NSF's ChemMatCARS, The University of Chicago, Lemont, IL 60439, United StatesPresent Addresses

**Notes**

The authors declare no competing financial interest.

**ACKNOWLEDGMENTS**

J.Z thanks Prof. Xutang Tao for providing valuable support and fruitful discussions. Work at Shandong University was supported by the National Natural Science Foundation of China (12374457 and 12074219), the TaiShan Scholars Program of Shandong Province (tsqn201909031), the QiLu Young Scholars Program of Shandong University, the Crystalline Materials and Industrialization Joint Innovation Laboratory of Shandong University and Shandong Institutes of Industrial Technology (Z1250020003), and the Project for Scientific Research Innovation Team of Young Scholars in Colleges and Universities of Shandong Province (2021KJ093). NSF's ChemMatCARS, Sector 15 at the Advanced Photon Source (APS), Argonne National Laboratory (ANL) is supported by the Divisions of Chemistry (CHE) and Materials Research (DMR), National Science Foundation, under grant number NSF/CHE-1834750. This research used resources of the Advanced Photon Source; a U.S. Department of Energy (DOE) Office of Science user facility operated for the DOE Office of Science by Argonne National Laboratory under Contract No. DE-AC02-06CH11357.